\documentstyle[preprint,aps]{revtex}
\draft

\begin{document}

\title{Dynamics of Bose condensed
gases in highly deformed traps}

\author{S. Stringari}
 \address{Dipartimento di Fisica, Universit\`{a} di Trento,}
\address{and Istituto Nazionale di Fisica della Materia,}
\address{I-3850 Povo, Italy}

 \date{December 30, 1997}

 \maketitle

 \begin{abstract}
We provide a unified  investigation of normal modes and sound propagation
at zero temperature in Bose condensed gases confined in highly asymmetric
harmonic traps and interacting with repulsive forces. By using hydrodynamic
theory for superfluids we obtain explicit analytic results for the
dispersion law of the low energy discretized modes  for both cigar and disk
shaped geometries, including the regime of large quantum numbers where
discrete modes can be identified with phonons. The correspondence  with
sound propagation in cylindrical traps   and the one-dimensional nature of
cigar type configurations are explicitly discussed.
 \end{abstract}

 \pacs{PACS numbers: 03.75.Fi, 05.30.Jp, 32.80.Pj, 67.90.+z}

\narrowtext

The recent experimental realization \cite{bec} of Bose-Einstein
condensation in dilute gases of alkali atoms confined in magnetic traps
has opened unique perspectives in the study of the dynamical and
statistical behaviour of mesoscopic quantum systems. These highly
degenerate quantum gases have been shown to exhibit both the dynamic
features of microscopic  many-body systems, associated with the 
occurrence of quantized collective excitations (normal modes),  
and the ones of macroscopic quantum fluids, characterized by 
the propagation of sound in the collisionless regime (zero sound).

The frequencies of the collective excitations have been the object of
various experimental  \cite{jila2,mit2,mit3} and theoretical studies,
 based on both analytic \cite{ss,fetter,castin,hung}  and
numerical \cite{oxford} investigations. The magnitude of these
frequencies is fixed  by the oscillator frequency of  the trapping
potential (typically a few ten Hz), the exact value depending on
the nature of the excitation (angular momentum etc.) and on the
effects of two-body interactions. In the experiment of \cite{mit3} the
large number of Bose condensed atoms and
the elongated geometry  of the trap have allowed for very precise
"in situ" images of the oscillations of the axial radius and consequently 
for high precision in the measurement of  the collective
frequencies, whose values turn out to be in excellent agreement with the
predictions of theory \cite{ss}. These  experiments, however, do not
probe directly the phononic nature of the excitation because of 
the discretization of frequencies imposed by the harmonic trapping.

The fact that phonons can propagate in the medium in a continuous way,
according to  Bogoliubov theory, has been recently
shown in the experiement of \cite{mit4} where wave packets have been
produced and directly observed in a Bose condensed gas of sodium  atoms
confined in a highly asymmetric, cigar type trap.

The purpose of this letter is to discuss the correspondence between these 
two dynamic  features (occurrence of discretized collective modes
and phonon-like propagation)  and to give analytic results for the
dispersion law of the collective modes in the case of  highly deformed
traps, including the regime of large quantum numbers where the discrete
modes can be identified with phonons.

The proper theory to describe the dynamic  behavior of interacting Bose
gases is the time dependent Gross-Pitaevskii \cite{GP} equation for the
order parameter. This equation can be written in a convenient form by
expressing the order parameter in terms of its modulus and phase,  $\Phi
=\sqrt{\rho}e^{i\phi}$, and looking for equations for the density $\rho$
and for the velocity field ${\bf v}=\mbox{\boldmath$\nabla$} \phi$. 
The equations of motion  then take the following form \cite{ss}:
\begin{equation}
 \frac{\partial}{\partial t} \rho + \mbox{\boldmath$\nabla$}
({\bf v}\rho) = 0
 \label{continuity}
 \end{equation}
and
\begin{equation}
m \frac{\partial}{\partial t} {\bf v} +
\mbox{\boldmath$\nabla$}( V_{ext} + g\rho -
\frac{\hbar^2}{2m \sqrt{\rho}}\nabla^2\sqrt{\rho} + \frac{1}{2}m{\bf
v}^2) = 0
\label{Euler}
 \end{equation}
where $g=\frac{4\pi \hbar^2 a}{m}$ is the interaction coupling constant,
fixed by the $s$-wave scattering length $a$ and
\begin{equation}
V_{ext} = {1\over2}m\omega_{\perp}^2r_{\perp}^2+{1\over 2}m\omega_z^2z^2
\label{vext}
\end{equation}
is the trapping potential, for which we  have made the choice  of  an
axially symmetric oscillator ($r_{\perp}=\sqrt{x^2+y^2}$ is the radial
coordinate). Eq.(\ref{continuity}) is the equation of continuity, while
(\ref{Euler}) establishes the irrotational nature of the motion. These
equations have  the typical structure  of the dynamic equations  of
superfluids at zero temperature (see for example \cite{NP}).

If the repulsive interaction among atoms is enough strong,
than the density profiles become smooth and one can safely
neglect the kinetic pressure term proportional to $\hbar^2$ in the equation
for the velocity field. This yields, for the static solution of
(\ref{Euler}),  the so called Thomas-Fermi expression
\begin{equation}
\rho_0({\bf r}) = \frac{1}{g}
\left ( \mu - V_{ext}({\bf r}) \right )
\label{tfgs}
\end{equation}
if $\mu\ge V_{ext}({\bf r})$, and zero elsewhere. Here $\mu$ is
the chemical potential, fixed by the normalization of $\rho({\bf r})$. In
the case of harmonic trapping one has 
\begin{equation}
\mu ={1\over 2}\hbar\omega_0(15N{a\over a_{HO}})^{2/5}
\label{mu}
\end{equation}
where $\omega_0 = (\omega_z\omega_{\perp}^2)^{1/3}$ is the geometrical
average of the oscillator frequencies  and $a_{HO}=
\sqrt{\hbar^2/m\omega_0}$ is the corresponding oscillator length. 
 The Thomas-Fermi
approximation (\ref{tfgs}) for the ground state
is accurate to the extent that the conditions $\mu \gg \hbar\omega_z ,
\hbar\omega_{\perp}$ are satisfied. The density (\ref{tfgs}) has the shape
of an ellipsoid with radial ($R_{\perp})$ and
axial ($Z$) radii defined by
$m\omega_{\perp}^2R^2_{\perp}=m\omega^2_zZ^2=2\mu$.

In the following we will neglect the kinetic energy pressure term also in 
the solution of the time dependent eqs. (\ref{continuity}, \ref{Euler}).
After linearization these equations then take the simplified,
hydrodynamic form
\begin{equation}
{\partial^2 \delta \rho\over \partial t^2} = 
\mbox{\boldmath$\nabla$}\left(c^2({\bf
r}) \mbox{\boldmath$\nabla$}\delta\rho\right)
\label{HD}
\end{equation}
where $\delta\rho = \rho({\bf r},t)-\rho_0({\bf r})$ and $c({\bf r})$,
defined by 
\begin{equation}
mc^2({\bf r}) = \mu - V_{ext}({\bf r}) ,
\label{c2}
\end{equation}
can be interpreted as a local sound velocity.  The validity of the
HD equations (\ref{HD},\ref{c2})   is based on the assumption that the
spatial variations of the density are smooth  not only in the 
ground state, but also during the oscillation.  In a uniform system
($V_{ext}=0$) this is equivalent to imposing that the
collective frequencies should be much smaller than  the chemical
potential. In this case  the solutions of the HD equation (\ref{HD})
are sound waves propagating with the Bogoliubov
velocity $c=\sqrt{\mu/m}$  where $\mu=g\rho_0$ and $\rho_0$ is the
equilibrium density. Due to the non uniform nature of the trapping
potential the solutions of (\ref{HD},\ref{c2})
exhibit new interesting features. The corresponding spectrum is
discretized and its explicit form,  in the case of  spherical trapping, was
first derived in \cite{ss}. The fact that these  equations have
analytic solutions  reflects the occurrence of underlying, non trivial
symmetries as recently discussed in \cite{hung}. For the lowest
multipolarities it is possible to obtain  analytic results  also
for arbitrary values of $\omega_z$ and $\omega_{\perp}$.
In particular, for the lowest $m=0$ value of the $z$-th component of
angular momentum and even parity, one finds \cite{ss}
\begin{equation}
\omega^2 (m=0) = 2\omega_{\perp}^2 + {3 \over 2} \omega_z^2 \mp {1\over
2}  \sqrt{9 \omega_z^4 -16
\omega_z^2\omega_{\perp}^2+16 \omega_{\perp}^4}
\label{pm}
\end{equation}
The dispersion (\ref{pm}) has been also derived in \cite{castin} using
different approaches.

In the limit of highly deformed, cigar shaped geometry
($\omega_z\ll \omega_{\perp}$), Eq.(\ref{pm})  gives the result
$\omega=\sqrt{5/2}\omega_z$   and $2\omega_{\perp}$ for
the low and high energy solutions respectively. As
anticipated in the introduction,  in the experiment of \cite{mit3},
carried out on a very asymmetric trap ($\omega_z/\omega_{\perp} =
17/230$), it has been possible to measure at low temperature the
frequency of the  low energy $m=0$ mode of even parity with very high
precision   ($\omega_{exp}=1.569(4)\omega_z$) in excellent agreement with 
theory ($\sqrt{5/2}=1.58$). This agreement is not a surprise because the
conditions of applicability of hydrodynamic theory are very well satisfied
in this experiment. In fact the parameters  of the trap ($a/a_{HO} \sim
10^{-3}$) and the number of atoms ($N \sim 10^7$) are such that
$\mu \sim 30\hbar \omega_{\perp} \sim 400\hbar \omega_z$ and hence the
validity of the Thomas-Fermi approximation is well ensured. Furthermore,
since the lowest frequencies are of order  $\omega_z$ the condition
$\hbar\omega\ll \mu$, relevant for the applicability of hydrodynamic
theory,  is also very  well satisfied.

It is useful to derive the dispersion law also for the
excitations with higher quantum numbers in cigar shaped traps. The
excitations we are interested in have  frequencies of order $\omega_z$,
much smaller than $\omega_{\perp}$. Let us rewrite the linear equations
(\ref{HD},\ref{c2}) in the form \begin{equation}
\omega^2\delta\rho=
- {1\over m}\nabla_{\perp}
\left((\mu-V_{ext})\nabla_{\perp}\delta\rho\right)
-{1\over m}\nabla_z\left((\mu-V_{ext})\nabla_z\delta\rho\right).
\label{HDHO}
\end{equation}
This equation shows that in the limit $\omega_z\ll \omega_{\perp}$ the low
energy  solutions  cannot have any dependence on the
radial coordinates. This would in fact yield high frequencies components of
order $\omega_{\perp}$ in the solution, due to the first term in
the r.h.s of (\ref{HDHO}). It is then natural
to expand the  $m=0$ solutions of (\ref{HDHO}) in the form
\begin{equation}
\delta\rho({\bf r}) = \delta\rho_0(z) +
\lambda^2r_{\perp}^2\delta\rho_1(z)
+ ...
\label{delta01}
\end{equation}
where $\lambda=\omega_z/\omega_{\perp}$ is the deformation parameter
of the trap. After inserting (\ref{delta01}) into  the hydrodynamic
equation (\ref{HDHO}) and integrating over  the radial coordinates
we obtain, for $\lambda\to 0$,  the following differential equation for
$\delta\rho_0$: \begin{equation} \omega^2\delta \rho_0(z) =
-\frac{1}{4}\omega_z^2(Z^2-z^2)\nabla^2_z\delta \rho_0(z)
+\omega_z^2z  \nabla_z\delta \rho_0(z) .
\label{HD1D}
\end{equation}
where $-Z\le z\le Z$. The discretized solutions of (\ref{HD1D}) are
polynomials  of the form  $\delta \rho^{(k)}_0(z) = (z^k +
\alpha z^{k-2} + ..)$, satisfying the orthogonality condition
$\int_{-Z}^{+Z}dz(Z^2-z^2)\delta\rho_0^{(k)}(z)
\delta\rho_0^{(k^{\prime})}(z)$ for $k\ne k^{\prime}$. They obey   the
dispersion relation   \begin{equation}
\omega^2 = {1\over 4} k(k+3) \omega_z^2 .
\label{kk+3}
\end{equation}
already derived in \cite{hung} using a different approach. The number of
nodes of these solutions  is equal to $k/2$ for even $k$, and to
$(k+1)/2$ for odd $k$. 

It is also interesting to   look for
solutions of (\ref{HD1D}) localized in the center of the trap
($z\sim 0$). These are sound waves propagating 
with velocity \begin{equation} c_{1D}=\sqrt{\mu/2m} \label{c1}
\end{equation}
where  we have used the identity $\mu={1\over2}m\omega_z^2Z^2$ for the
chemical potential. Notice that in the Thomas-Fermi approximation the
chemical potential is  always related to the central density  by the
relation $\mu = g\rho_0(0)$ (see eq.(\ref{tfgs})), so that the sound
velocity $c_{1D}$ is smaller  by a factor $\sqrt2$ with respect to the
Bogoliubov velocity calculated in the center of the trap. The occurrence of
this factor was first pointed out in \cite{zaremba}  and has a simple
physical meaning. In fact, in deriving the relevant hydrodynamic 
eq.(\ref{HD1D}), we have integrated (\ref{HDHO})
over  the radial variables, so that the new sound
velocity corresponds to an average whose value is  smaller than the one in
the center of the trap. 

To better understand the propagation of sound waves  in the case of
highly elongated traps let us consider  a trap with cylindrical geometry
 and harmonic confinement in the radial direction. The hydrodynamic
equations in this case are simply obtained by setting $\omega_z=0$ in
(\ref{vext}) and take the form
\begin{equation}
\omega^2\delta\rho =
-\frac{1}{2}\omega_{\perp}^2\nabla_{\perp}\left((R_{\perp}^2-
r_{\perp}^2) \nabla_{\perp}\delta
\rho\right)
-\frac{1}{2}\omega_{\perp}^2(R_{\perp}^2-r_{\perp}^2)\nabla^2_z\delta
\rho
\label{HDC}
\end{equation}
defined in the interval $-L<z<L$,  where $2L$ is the length of the
cylinder, and $0<r_{\perp}<R_{\perp}$. It is worth noting that in the
cylindrical geometry the validity of the Thomas-Fermi approximation  for
the ground state is guaranteed by the condition $\mu \gg
\hbar\omega_{\perp}$ or, equivalently, $Na/L\gg 1$ (in this
case we always assume $L\gg R$).  If we impose periodic boundary conditions
 at $z=\pm L$ the solutions of (\ref{HDC})
can be written in the form  $\delta\rho=(\delta\rho_0(z) + r_{\perp}^2
\delta\rho_1(z)+..)$ with $\delta\rho_0$ and $\delta\rho_1$ proportional
to $e^{iqz}$. After integration  in the radial
variables eq.(\ref{HDC}) takes, to the lowest order in $q^2R_{\perp}^2$,
the simplified form $\omega^2\delta\rho_0=
-{\mu\over2m}\nabla^2_z\delta\rho_0$, yielding the dispersion
$\omega=c_{1D}q$  with the sound velocity given by (\ref{c1}). The
numerical solution of (\ref{HDC}) with
larger $q$  has been carried out in \cite{zaremba}). It is not difficult
to calcualate the first correction to the linear behavior. One finds 
$\delta\rho_1(z) = q^2\delta\rho_0(z)/8$ and
\begin{equation}
\omega^2 = c_{1D}^2 q^2 (1- {1\over48} q^2R_{\perp}^2)
\label{c1q}
\end{equation}
Result (\ref{c1q}) explicitly reveals that the linear  dispersion holds
if the wavelength  is much larger than the radial size of the 
condensate and that the sound has a negative dispersion \cite{zaremba}.

Coming to the dynamic behavior in the presence of harmonic trapping
one expects to observe wave packets propagating with the $1D$ sound
velocity  $c_{1D}$ if the conditions $qZ\gg 1$ and $qR_{\perp} \ll 1$
are simulataneously satisified. Of course the condition $\hbar q \ll  mc$
must be also satisfied because it ensures the applicability of
hydrodynamic theory. The  condition $qZ\gg 1$
guarantees that the medium can be treated as  locally uniform in the $z$
direction and that one can consequently observe wave packets propagating
in the central region of the trap.  The  condition $qR_{\perp} \ll 1$
instead ensures that we are not exciting the
motion in the  radial direction and that the dispersion will look  "one
dimensional" and given by the first term of (\ref{c1q}). In the
experiment of \cite{mit4}  $Z$ is a few hundred microns,
$R_{\perp}=\lambda Z$ is a few ten microns   and
$mc/\hbar \sim 2-4 (\mu m)^{-1}$ depending on the value of the peak
density. It is then possible that the wave packets observed in
\cite{mit4} are, at least partially, built up with
wave  vectors satisfying the above conditions. A detailed discussion of
the propagation of wave packets, with the inclusion of non linearity
effects,  has been recently reported in
\cite{pethick}

Let us complete our discussion by calculating  the first corrections to 
the dispersion relation (\ref{kk+3}). By solving the HD equations
(\ref{HDC}) to the next order in $\lambda^2$ we obtain, after some
straightforward algebra, the result
$\delta\rho_1(z) = -{1\over 8}\nabla^2_z\delta\rho_0(z) $  (notice the
analogy with the result $\delta\rho_1 = q^2 \delta\rho_0/8$ holding in
 cylindrical geometry) and
\begin{equation}
\omega^2 = {1\over 4} k(k+3)\omega_z^2\left(1-{\lambda^2\over
48}(k-1)(k+4)\right) .
\label{omega24}
\end{equation}
Some remarks are in order here. First one recovers the limit of the
sound wave dispersion (\ref{c1q}) holding in cylindrical geometry
in the limit of large quantum numbers $k\gg 1$, by the proper identification
\begin{equation}
c^2_{1D}q^2 = k^2\omega_z^2/4
\label{c1id}
\end{equation}
yielding $k^2 = q^2Z^2$ . This is consistent with the already discussed
condition $qZ \gg 1$ needed to observe phonons propagating in the
$z$-th direction. In the same limit the first corrections in the dispersion
(\ref{c1q}) and (\ref{omega24}) coincide since one has $q^2R_{\perp}^2 =
k^2\lambda^2$. This completes the correspondence  between the propagation
of discretized modes and  sound  and
shows the analogy between the dynamic behavior in the cylindrical and
elongated harmonic oscillator geometries.

Concerning the frequencies of the discretized modes  predicetd by
(\ref{omega24}) it is worth pointing out that the lowest mode $(k=1)$
corresponds to the center of mass motion and its
frequency coincides  with the oscillator frequency $\omega_z$. This
frequency is unaffected by the presence of two-body interactions. The
second mode ($k=2$)  is the "quadrupole" collective excitation observed
in \cite{mit3}. It corresponds to the low energy solution  (\ref{pm})
 in the $\lambda \to 0$ limit. The direct experimental observation of 
the higher modes, as well as of the first 
correction in $\lambda^2$ predicted by
(\ref{omega24}) would complete the scenario of the low energy  
excitations in the elongated geometry.

It is finally interesting to discuss the one-dimensional nature of these
systems. One should first point out that all the results discussed in
this paper have been derived starting from 3D configurations. In
particular, in order to derive the dispersion law
(\ref{omega24}), we have assumed the validity of the Thomas Fermi
approximation in both axial and radial  directions, so that
(\ref{omega24}) holds if $\mu \gg \hbar \omega_{\perp} \gg
\hbar\omega_z$. This means that the ground state wave function of the
system is built up including many excited single particle states in both
axial and radial directions. A full 1D problem  should involve only the
lowest oscillator wave function in the radial direction and in this case the
corresponding excitation spectrum in the Thomas-Fermi regime would be
\cite{ho}  $\omega^2 = {1\over 2}k(k+1)\omega_z^2$ instead of
(\ref{omega24}). This dispersion is easily derived from (\ref{HDHO})
ignoring the radial coordinates in the equation and holds if 
$\hbar\omega_{\perp} \gg (\mu-\hbar\omega_{\perp})\gg \hbar\omega_z$. At
present this regime is  far from experimental possibilities. Nevertheless,
even remaining in the 3D Thomas-Fermi regime, it is clear that for highly 
elongated configurations the low energy dynamic behavior ($qR_{\perp} \ll
1$)  looks one-dimensional, the radial  directions providing only a
renormalization of the sound velocity. So all the statistical and
thermodynamic properties of 1D systems should apply to  these
configurations provided the temperature is smaller than the radial
oscillator  energy. This includes in particular the Luttinger liquid-like
behavior recently suggested for these systems  \cite{lutt} and the
two-step Bose-Einstein condensation recently discussed  in
the context of the ideal Bose gas \cite{druten}. Furthermore, due to the coupling with
the radial modes, the transition from the phonon to the  single particle
regime exhibits new interesting features. In particular the first correction
to the phonon dispersion is negative (see (\ref{c1q})  and
(\ref{omega24})), differently from the traditional Bogoliubov behavior 
$\omega^2=c^2q^2 +(q^2/2m)^2$.

In analogous way we can carry out the analysis in the disk geometry
($\omega_z\gg \omega_{\perp}$). In this case, to the lowest order in
$1/\lambda^2$, the density fluctuations will depend only on the radial
coordinates and, after integration  of Eq.(\ref{HDHO}) in the variable 
$z$, the relevant equation for $\delta\rho({\bf r_{\perp}})$ takes the form
\begin{equation}
\omega^2\delta\rho({\bf r_{\perp}}) =
-\frac{1}{3}\omega_{\perp}^2
((R_{\perp}^2-r_{\perp}^2)\nabla^2_{\perp}\delta
\rho({\bf r_{\perp}})
+\omega_{\perp}^2{\bf r_{\perp} \cdot \nabla_{\perp}}\delta \rho({\bf
r_{\perp}})
\label{HD2D}
\end{equation}
Notice that in this case wave packets in the center of the trap
will propagate  with the "2D" sound velocity
\begin{equation}
c_{2D} = \sqrt{{2\over 3}{\mu\over m}} .
\label{c2D}
\end{equation}
The discretized solutions of (\ref{HD2D})) have the form
\begin{equation}
\delta\rho^{(n,m)} = (r_{\perp}^{2n} + \alpha r_{\perp}^{2n-2}
+..)r_{\perp}^m e^{im\phi}
\label{delta2d}
\end{equation}
where $m$ is the $z$-th component of  angular momentum  and $n$ fixes the 
number of radial nodes. They satisfy the orthogonality condition
$\int_{r_{\perp}\le R_{\perp}} d{\bf r}_{\perp}
(R^2_{\perp}-r^2_{\perp})^{1/2}
\delta\rho^{(n,m)}\delta\rho^{(n^{\prime},m^{\prime})} = 0$ for $n,m \ne
n^{\prime},m^{\prime}$. The resulting dispersion takes the form
\begin{equation} \omega^2 =({4\over 3}n^2+{4\over 3}nm +2n +m)
\omega_{\perp}^2. \label{dispersion2d} \end{equation} The cases $n=0, m=2$
($\omega=2\omega_{\perp}$), and $n=1, m=0$
($\omega=\sqrt{10/3}\omega_{\perp}$) corespond to the modes of even parity
 observed in the experiments of \cite{jila2} (the $m=0$ state corresponds 
to the low energy solution  (\ref{pm}) in the $\omega_{\perp}\ll
\omega_z$ limit). One should however note that
in this experiment the deformation of the trap  and the
number of atoms in the condensate  were  not very large ($\lambda
=\sqrt8$ and $N\sim 10^4$). As a  consequence the conditions required to
apply the dispersion  law (\ref{dispersion2d}) ($\lambda\gg 1$ and
validity of the Thomas-Fermi approximation) are not very well satisfied
in this case.

We finally note that also in  deriving the dispersion law
(\ref{dispersion2d}) we have assumed  the validity of the Thomas-Fermi
approximation in both axial and radial directions
($\mu \gg \hbar\omega_z \gg \hbar\omega_{\perp}$). The hydrodynamic
dispersion in a true 2D trap would in fact follow a different
dispersion law given by \cite{ho} $\omega^2 = (2n^2 + 2nm +2n
+m)\omega^2_{\perp}$. This 2D hydrodynamic dispersion law holds  if the
conditions $\hbar\omega_z\gg(\mu- {1\over 2}\hbar\omega_z)\gg \hbar
\omega_{\perp}$ are satisfied.

\bigskip 

Stimulating discussions with Dan M. Stamper-Kurn and Wolfgang Ketterle are
acknoweledged.

\end{document}